# Transdisciplinary AI Education: The Confluence of Curricular and Community Needs in the Instruction of Artificial Intelligence

Roozbeh Aliabadi[1(✉)], Aditi Singh[2(✉)], and Eryka Wilson[1,3]

[1] ReadyAI, Pittsburgh, USA
{rooz,eryka}@readyai.org
[2] Cleveland State University, Cleveland, USA
a.singh22@csuohio.edu
[3] Keiser University, Fort Lauderdale, USA

**Abstract.** The integration of artificial intelligence (AI) into education has the potential to transform the way we learn and teach. In this paper, we examine the current state of AI in education and explore the potential benefits and challenges of incorporating this technology into the classroom. The approaches currently available for AI education often present students with experiences only focusing on discrete computer science concepts agnostic to a larger curriculum. However, teaching AI must not be siloed or interdisciplinary. Rather, AI instruction ought to be transdisciplinary, including connections to the broad curriculum and community in which students are learning. This paper delves into the AI program currently in development for Neom Community School and the larger Education, Research, and Innovation Sector in Neom, Saudi Arabia's new megacity under development. In this program, AI is both taught as a subject and to learn other subjects within the curriculum through the school system's International Baccalaureate (IB) approach, which deploys learning through "Units of Inquiry." This approach to education connects subjects across a curriculum under one major guiding question at a time. The proposed method offers a meaningful approach to introducing AI to students throughout these Units of Inquiry, as it shifts AI from a subject that students "like" or "not like" to a subject that is taught throughout the curriculum.

## 1 Introduction

Artificial Intelligence is an exponentially growing technology aimed at improving and reshaping our daily lives. The future lies in AI-infused society, and it is schools' responsibility to prepare young minds including kindergarten to eighth grade to thrive in a rapidly transforming society. Young people may need to understand AI basics to flourish in science and engineering. Hence, developing an AI curriculum at elementary school or high school becomes a necessity to promote students' willingness and bolster their intention to understand AI. AI education for students should emphasize learning contents on hands-on AI activities rather than theory. The adoption of advanced technologies to provide students' hands-on experience is a trend in ML teaching [1].

The education research community agrees that school curriculum must transition from traditional expositive classes to more casual, interactive, experimental, and collaborative learning environments. The most important trend in this direction is the experience learning theory [2, 3]. The iSTREAM and iCDIOS AI teaching concepts were first put forth by Liu et al. [4] at the American Engineering Education Conference ASEE. Incorporating STEM (Science, Technology, Engineering, and Mathematics) and CDIO (Integrated Teaching Methods of Conceiving, Designing, Implementing, and Operating for Universities) from the US and Europe [4].

Educators have introduced AI literacy in school at different levels, such as elementary or high school. AI curriculums should educate future generation by introducing AI. However, the following question remains unanswered: Would a curriculum improve students' attitude towards ac-quiring a knowledge of AI? Will students understand a global perspective of AI? Are the current curriculums inclusive of the teaching AI in such a way that students can understand the process of AI and impact of AI? To facilitate answering these, an investigation of the current AI curriculum being offered is first necessary. An investigation of current AI curriculum being offered demonstrates that the answer to these questions is no and that a new approach of transdisciplinary AI education is needed.

## 2 Related Works

### 2.1 Current AI Curriculums

Current AI curriculums are mostly standalone educational content with the focus of introducing AI as a topic in schools. Introducing the topic does not make young generation AI ready. Educators developed a curriculum with a focus on "why and what to teach" to cultivate students' AI literacy. Their focus was to introduce the topic based on three levels: AI knowledge, AI skills, and AI attitude. The modules varied from sub tasks like face detection, image classification, text classification and self-driving cars [5]. It is important for the students to know each subtask and their impact on society. Unfortunately, teaching about AI concepts is siloed and leads to faulty assumptions. Students should not think AI as just coding, but also comprehend the basics.

Lee created a middle-school AI literacy curriculum on ethics education. This was an attempt to investigate children's AI literacy through K12 education. It incorporates algorithmic bias, ethical recommender system design, and unanticipated consequences into a variety of AI learning activities like using Teachable Machine to train models, creating texts, and redesigning videos [6].

### 2.2 Transdisciplinary AI Education

Transdisciplinary AI education is an approach to teaching and learning about Artificial Intelligence (AI) that incorporates multiple disciplines and perspectives in order to provide a comprehensive and well-rounded understanding of the field. It acknowledges that AI is a complex and interdisciplinary field that draws on concepts and techniques from computer science, mathematics, psychology, linguistics, philosophy, and many other fields. In a transdisciplinary AI education program, students may have the opportunity to learn about AI from a variety of angles and contexts, rather than focusing solely on technical concepts and techniques. This may include exploring the ethical and social implications of AI, considering the potential impacts of AI on different industries and fields, and learning about the ways in which humans and AI systems can work together effectively.

Researchers [7] discussed the potential role of AI in transdisciplinary education, including the benefits of using AI to bring together multiple disciplines and the challenges that educators may face when incorporating AI into the curriculum. Researchers [8] discussed the current state of research on transdisciplinary AI education for middle school students, including the benefits of incorporating multiple disciplines into the curriculum and the challenges that educators may face. It also provides recommendations for how to effectively teach these topics to middle school students. One key component of transdisciplinary AI education is multidisciplinary coursework, which involves the integration of multiple disciplines in a single course or program. This approach allows students to gain a more holistic understanding of AI and its applications, as well as develop skills in areas such as data analysis and machine learning. Several studies have demonstrated the effectiveness of multidisciplinary coursework in AI education [9–11].

Another important aspect of transdisciplinary AI education is experiential learning, which involves hands-on learning experiences such as internships, capstone projects, and industry partnerships. This approach allows students to apply their knowledge and skills in real-world settings and gain valuable industry experience. Studies have shown that experiential learning can improve student outcomes and increase the attractiveness of AI programs to industry partners [11, 12]. The goal of transdisciplinary AI education is to provide a broad and comprehensive understanding of the field that goes beyond technical expertise and considers the wider social and cultural context in which AI is being developed and used.

### 2.3 Project Based Learning in AI

Focusing on the convergence of innovative technology, activities, and approaches to include students in creative learning experiences, the contemporary framework on constructionism in learning developed by MIT Media Lab. The framework is based on four basic principles (the "Four P's of Creative Learning") that encourage students to be-come creative learners. Project being the first principle, where the researcher discovered students learn best when they actively participate in the significant initiatives that interest them. The second principle is passion, where the flexibility to work on meaningful projects encourages students to invest in their work and persevere in the face of adversity, allowing them to learn more in the process. The third principle is Peers in which learning is considered as a social activity and students frequently share ideas and collaborate on projects. The fourth principle is play, which encourages learning via fun experimentation, trying new things, fiddling with tools and materials, pushing limits, taking calculated risks, and repeatedly iterating projects [13].

Project based learning in classrooms not only incorporates engagement and interaction but also provides students an opportunity to acquire deeper knowledge and meaningful skills. To keep up with the pace of such a digital evolutionary process and ensure that future generations immerses in this new technological context, secondary, technical, and higher learning programs have increased the frequency and accessibility of teaching basics of AI to all levels of society [14, 15]. Similar studies devoted to the development of AI learning content combined with low-cost robots [16].

Researchers are only now focusing on how young people can design and create ML applications. The few examples provided describe ML projects and workshops where children create models of their own physical activity, investigate object recognition through their own drawings, and envision futuristic smart toys and gadgets [17–20].

Burgsteiner [21] explored the effectiveness of AI literacy in high school by developing iRobot, an AI-course for teaching students' important topics of AI/computer science by combining theory and lab. Students could communicate with the robot via voice commands and interactive tools. Following the initial play and interaction, the children were driven to code the agents via coding apps. The coding apps used in this study were based on Scratch and visual block programming language.

### 2.4 Problem Based Learning in AI

Researchers hypothesized that; besides project-based learning, the future of STEM education must integrate imaginative creativity and embed in a dynamic social and cultural environment to prepare students for the complex and unpredictable futures [22]. In problem-based learning, students learn to solve complex and realistic problem or an area by working in collaborative groups under supervision of faculty.

In [23] researchers discusses an educational program, PRIMARY AI that uses AI tools to teach students about AI concepts and techniques through problem-solving activities in the context of life sciences. The authors argue that PRIMARY AI is an effective approach to teaching AI, as it allows students to learn through inquiry-based adventures and experience the power of AI firsthand. They also highlight the benefits of using block-based programming in PRIMARY AI, as it allows students to easily learn and apply AI concepts and techniques. The authors present the results of a pilot study in which they tested PRIMARY AI with a group of middle school students and found that the program was successful in increasing student engagement and understanding of AI. The authors also discuss the challenges and opportunities of implementing PRIMARY AI in education and provide recommendations for educators interested in using this approach.

Researchers have suggested methods for integrating artificial intelligence with social and real-world issues in AI education by putting the concept of "AI challenges" into practice to inspire students to learn about AI [24]. Researchers have also combined other subjects, such as airport security, wildlife preservation, and other social issues, to develop AI challenges. This technique helps in the decentralization and dissemination of AI knowledge to a variety of audiences, including public high school students, university undergraduates, law enforcement personnel, and even park rangers who preserve animals [25, 26].

In [27], Van Brummelen and Lin (2021) discuss the development of an integrated AI curriculum for K-12 classrooms, with a focus on engaging teachers in the co-design process. The authors argue that involving teachers in the design process is critical for the success of AI education initiatives, as it ensures that the curriculum is relevant and practical for teachers to implement in their classrooms. The authors present the results of a case study in which they worked with a group of teachers to co-design an AI curriculum and found that the teachers valued the opportunity to collaborate and share their expertise. The authors also discuss the challenges and opportunities of implementing the curriculum in K-12 classrooms and provide recommendations for educators interested in using this approach. However, the authors do not provide data on student outcomes or the effectiveness of the curriculum, which makes it difficult to assess the impact of the co-design process on student learning. Further research with a larger sample and more robust data on student outcomes would be needed to understand the effectiveness of this approach more fully.

## 3 Implementation of a Wholistic AI Curriculum Neom, Saudi Arabia Curriculum

The AI curriculum created for Neom Community School (NCS) included both project-based and problem-based learning exercises for students at the school, which included kindergarten through eighth grade. Moreover, this curriculum integrated such learning methodologies by introducing both "challenge" assignments and "mini-projects" throughout the design process. Such "challenge" assignments occurred during the middle of a lesson, presenting students with a challenge mid-lesson. For instance, one lesson might begin by presenting content to students. After some direct instruction, we presented students with a specific instance where they need to apply the direct instruction or where they need to solve a problem in preexisting code. By introducing such challenges, we encouraged students to work in teams, collaborate and communicate to present ideas to address the challenge, think critically about the instruction presented to them, analyze the issue at hand, and generate novel solutions that perhaps even the instructor did not envision during the design process. The fifth lesson in Neom's Beginner AI & Calypso Curriculum acknowledge such a challenge. After learning about AI motion, object recognition, and object interaction, we presented students with a challenge that requires them to code the AI robot to navigate between objects and "put fires out". Thus, we instructed students to imagine a situation where AI was doing the firefighting, helping to negate putting humans at risk. We have given students the tools through coding instruction and other practice sessions with troubleshooting, but when faced with a challenge, we asked them to apply the information they have learned and their engagement up to that point.

This example also illustrates the importance of project-based learning. Students in this case are given a project to complete, to turn the AI robot into a firefighting entity. This project requires students to imagine a situation where AI could be useful to human society, and then to apply their knowledge of AI to create a project that shows the capacity of AI to help solve human problems. Later in the curriculum, students are given the opportunity to create their own projects, identifying areas of their concern that facilitate human AI interaction and positive societal impact thanks to the development of AI.

Besides challenges representing real-world situations, we present students with mini projects. These mini projects also meet the needs of learners engaged in problem-based learning and project-based learning. For instance, students are presented with a mini project towards the end of the Intermediate AI & Calypso Curriculum where students are, across multiple days, engaged in the ideation, creation, and presentation of a museum. Students identify the theme of the museum by combining their interests and the history of their individual cultures, given that the student body in Neom represents all parts of the world—from Saudi Arabia to South America to the South Pacific. Students then research pieces of art or other cultural or historical artifacts that the museum will showcase. After physically creating the museum, coding the actions of the AI robot, troubleshooting the challenges, and testing their projects, students prepare an outline and present their project. In this case, Neom Community School (NCS) hosted an AI night and invited parents of students engaged in the curriculum. Students then presented their projects to a large group of their peers, parents, teachers, and the broader Neom community.

Besides embodying project-based learning, this demonstrates problem-based learning. To start the project, the teacher discusses challenges museums face, including the expense of hiring tour guides, the inability of humans to remember the quantity of details necessary to speak cogently on thousands of pieces of artwork and artifacts of a museum, and tracking the inventory when artwork or artifacts are rotated, an issue commonly faced by museums across the world. Students propose manners that AI and AI robots could help address these problems, and they are encouraged to redress these problems within the curriculum. Students face discrete problems while coding their projects, and with the aid of a well-trained instructor, they are assisted—not told—how to overcome those problems in designing and coding their projects.

Finally, problem-based learning is illustrated in the limitations of AI. The AI robot will only respond to users according to its coding. For instance, when introducing itself, if only one line of coding is provided for the AI robot, then the robot will din that same line whenever it introduces itself. To museum guests who visit the museum several times, this may become monotonous show how AI does not meet the goal of Big Idea #4, i.e., Natural Interaction [28]. Thus, students are introduced to strategies to address this problem and they propose, test, and implement coding that ensures the AI robot does not seem hampered by the limitations of basic AI.

In all these facets, the curriculum written for and piloted by NCS shows how the implementation of AI instruction must address a range of needs, including challenging students with problems, facilitating their generation and creation—and presentation—or projects, and foster critical thinking in line with admirable goals as articulated by the vision of Neom.

### 3.1 AI Inquiry Curriculum

NCS also sought to present students with a more inquiry-based curriculum of AI. Thus, the "AI Inquiry Curriculum" was created, where students were introduced to a more independent presentation of AI concepts. From computer vision to machine learning, students inquire into the everyday uses of AI that have so surreptitiously and seamlessly integrated their way into people's everyday lives. In this series of discrete lessons, students were introduced to an example of AI in their lives. Then, they are taught to ask questions into its functionality and implications. Finally, each lesson concludes with students experimenting and inquiring into future uses and ramifications of the technology.

To use one example, the AI Inquiry Curriculum explores the use of facial recognition technology already used in Neom. Upon entering the Neom construction area, users' faces are scanned by a system. At which point, they are either greeted or asked to wait for a security guard who would then check for identification and permission to enter the facilities. To a computer scientist, this represents an obvious use of AI. However, to the general residents and visitors of Neom, this seems like just a quotidian experience. Students, who have become so acclimated to the process that they do not question the nature of the experience, are asked in one lesson of the AI Inquiry Curriculum to describe what technology is being used. By pausing for a moment and being prompted to think about what is taking place, students recognize that technology is being implemented seamlessly into their lives.

Students then move from recognizing the everyday and even subtle uses of technology to facilitate human experience. When asked what would happen if the AI technology were not used at the entry points to Neom, students recognize long queues might develop, which in Saudi Arabia might mean residents and visitors bake in the Arabian desert sun. Students are also prompted to inquire into how the technology works, even testing similar technology using black box apps. First, students are prompted to discover the power of the technology by obfuscating part of their face. At what point does the AI system no longer recognize a person? Does it require obscuring the lower part of the face or the upper part of the face? What if only half of the face is visible? By posing such questions, students are exploring the large limitations of the system. After training a system on a student's individual face, students then explore the limitations not just the computer vision and recognition tool but even machine learning and facial recognition. Students are prompted to wear glasses, hats, fake bears, mustaches, and so forth to see if the AI systems continuously recognize them first as a human being and second as the individual on which the system was trained. Finally, and true to the larger goals of project-based learning, students explore how such AI tools could be used to facilitate their own lives and the discrete needs of their communities.

This example represents just one lesson of the AI Inquiry Curriculum. Future expansion of the AI Inquiry Curriculum might explore through the inquiry into chat bots; user-experience recommendations such as those demonstrated by recommended videos on YouTube, Netflix, and Spotify; and AI generation of human-like writing, such as in the form of poems.

### 3.2 Benefits of AI Inquiry Curriculum

AI Inquiry in the manner presented above presents many benefits. For one, the lessons are non-sequential. This type of curriculum presents students with opportunities to engage a variety of material, some of which will not interest them as much as other material. Any reasonable instructor or curriculum designer will recognize that even the best lessons only reach a majority of students, not all.

Moreover, the benefits of this curriculum include introducing students to a wide range of real-life applications of AI. Because each lesson introduces students to an extant technology, students are prompted to reflect on how AI has already integrated itself into their lives. Additionally, through inquiry, students are asked to reflect on how such technology is not inherently agnostic. AI is not unimpactful in human life. Every bit of AI technology comes with societal impact, and no technology is inherently beneficial nor inherently deleterious. To use the example above, facial recognition facilitates the entry to the Neom facility, speeding up entry. However, such technology also allows the nascent city to employ fewer security personnel. It also documents and tracks individuals' location and entry to the site, raising questions of security and privacy.

Another interesting benefit is the flexibility this type of curriculum presents. As new technology finds its way into the marketplace and individuals' everyday lives, this approach allows lessons to be added, removed, and rethought entirely in order to present students with the most current curriculum possible. To use the example above, new examples of facial recognition apps are being introduced all the time as companies seek to release the next greatest app. These can be introduced at will, without having to redesign an entire program. It also allows students to challenge technology that is marketed as the next greatest app, questioning whether it in fact is or whether it should be universally adopted, based on its human interaction capabilities, its ability to collect users' personal data, and a cost-benefit analysis of its societal impact.

Finally, the method of inquiry into AI concepts prompts students to explore the uses of technology in a discrete manner. In other words, students are free to explore that technology for the lesson's sake. Students are not necessarily required to make larger connections to other curricular concepts. Students can learn about the technology for the technology's sake. This approach presents a certain freedom to explore a concept without having to explore larger connections of it, presenting a less stressful method of learning.

### 3.3 Challenges of AI Inquiry Curriculum

In their paper, Zhou et al. (2020) review existing AI curricula and identify common challenges and opportunities in teaching AI. The authors argue that there is a need for more comprehensive and practical AI curricula that are accessible to a wide range of learners. They also highlight the importance of involving educators in the development process and ensuring that the curriculum is relevant to the needs of students and teachers. The authors present a set of recommendations for designing effective AI curricula, including the use of real-world problems and case studies, the integration of multiple disciplines, and the incorporation of hands-on activities and projects [29].

When other institutions and well-meaning AI lesson designers present AI curriculum, even those with substantial reputation within the field, the lessons are disconnected and do not build towards a cohesive whole. While this design process allows for the aforementioned benefits, it also challenges the ability of designers to have larger objectives in a curriculum and measure whether students are meeting those objectives. When discrete lessons can only be grouped under a larger goal such as "Students will explore the use of current technology and analyze its impact," it becomes difficult to determine whether students will explore two or twenty current technologies.

Overall, when students learn about AI in isolation through a curriculum similar to the AI Inquiry one designed and implemented for Neom, their perspective and understanding of AI may be limited. They become analyzers, which is beneficial to an extent, but they do not become active creators and contribute to an AI ecosystem, where they can begin to recognize their role in designing their own AI-powered future. On the other hand, when paired with other curricula that introduce AI in a more sequential manner, as well as when AI is integrated in a transdisciplinary way, such as what is beginning to take shape at NCS, the benefits of exploring a broad base of technologies in the form of the AI Inquiry Curriculum become more salient. Designers of AI curricula would greatly benefit from an approach that pairs the sort of "one-off" lessons currently being produced with a more serial approach that builds towards larger and measurable goals, generating nascent AI designers and thinkers equipped to shape their communities and their world with the potential AI offers.

### 3.4 Calypso-Based Learning with an AI Robot

The Calypso-based curricula implemented software paired with the robot Cozmo. Cozmo is an AI-powered robot with the ability to engage AI to navigate, recognize faces, recognize objects, and interact with those objects. Thus, using Calypso, a non-sequential block-based coding software, students can learn about AI and major concepts in computer science.

The Calypso-based curricula were broken into three nine-week units consisting of one hour of material per lesson. In this first of three sequential programs, Beginner AI & Calypso Curriculum, students are introduced to Calypso, produced by Visionary Machines, LLC., a Pittsburgh-based company founded by Professor David Touretzky, a research professor at Carnegie Mellon University. In order to understand AI and how computer science governs the actions of AI, students are introduced to the governing laws of Calypso and how Cozmo interacts with the world through them and because of them [30]. Throughout the curriculum, students will demonstrate their learning in project-based exercises integrated into each lesson and culminating in a larger final project.

In the second of three sequential programs, Intermediate AI & Calypso Curriculum, students continue their exploration of AI through Cozmo and Calypso. In this curriculum, students learn about key concepts in Artificial Intelligence and computer science more broadly, including state machines, computer logic, and human-introduced randomness. Students also complete their first major project, a Museum Challenge, where students are asked to program Cozmo to navigate around a museum of students' creation, showcasing the artwork and interacting with visiting patrons.

This final project allows students to connect their own discrete cultures and interests with the functionality and artificial intelligence that Calypso and Cozmo permit students to engage. In the final of three sequential programs, Advanced AI & Calypso Curriculum, students focus their learning through PBL (Project Based Learning). They will complete several projects and learn about the highest order functionality in Calypso, including Teachable Machine and the use of walls. They will complete their learning about Cozmo through a curricular capstone project. Table 1 provides a programmatic overview of each along with overall student learning objectives and a topical breakdown.

**Table 1.** Curriculum with student learning outcome and topical breakdowns.

|  | Student Learning Outcome | Topics |
| --- | --- | --- |
| Beginner AI & Calypso Curriculum | • Identify the five laws of Calypso<br>• Compare and contrast actions based on computational thinking<br>• Create projects that demonstrate their understanding of the 5 Big Ideas in AI | 1. Introduction to Cozmo and Computer Vision<br>2. The First Law of Calypso<br>3. Motion with Cozmo<br>4. The Second Law of Calypso<br>5. Fireman Challenge - Project Based Learning Exercise<br>6. Multiple Actions and Conflict Resolution<br>7. The Third Law of Calypso<br>8. Dependent Actions & Rule Indentations<br>9. Introduction to State Machines |
| Intermediate AI & Calypso Curriculum | • Explain the function and use of state machines<br>• Distinguish between AI actions that use the "Not" function and other forms of logic<br>• Create and present a project that combines the beginning and intermediate functions of Calypso and Cozmo | 1. State Machines - Unplugged<br>2. State Machines & Dialogs, Timers, and Conditions<br>3. Multiple Characters in Programming<br>4. Negation & Introduction to Logic in Computer Science<br>5. The Museum Challenge - Part I<br>6. The Museum Challenge - Part II<br>7. The Museum Challenge - Part III & Presentations<br>8. Randomness in Calypso<br>9. Calypso and Audio Importations |

(*continued*)

**Table 1.** (*continued*)

| | Student Learning Outcome | Topics |
|---|---|---|
| Advanced AI & Calypso Curriculum | • Demonstrate their ability to build rooms and walls in Calypso<br>• Ideate a major Capstone Project<br>• Create, Troubleshoot, and Present their final Capstone Project | 1. Building Cozmo's Shack<br>2. Working with Walls and Rooms<br>3. Custom Map Layouts<br>4. Cozmo - The Home Aide Challenge<br>5. Teachable Machine and Calypso<br>6. Dogs vs. Rhinos – Differentiating Using a Visual Classifier and Cozmo<br>7. Calypso Curriculum Capstone I<br>8. Calypso Curriculum Capstone II |

### 3.5 Calypso-Based Programming and Problem-Based and Project Based Learning

Calypso programming presents students with unique challenges. Coding must be accurate and respect the laws of Calypso that govern Cozmo's thinking. Students are taught, like with any programming language, the need to troubleshoot and identify problems. Moreover, because of the nature of using Cozmo, students then get to see the problems in real life. Whereas perhaps a Scratch or Python program won't run, using the robot in real life allows student to see how Cozmo may begin to act erratically, or the visible actions are counter to what was expected in the coding, presenting unique challenges but also unique avenues, compared to most coding curricula, to experience a true problem-based curriculum.

Additionally, because of Calypso and the real-life manipulation of an AI powered robot, students can devise, create, code, and present real-life projects. From teaching Cozmo to navigate from cube to cube, putting out 'fires' as signified by glowing cubes, to having Cozmo navigate around a set representing a museum, a home, a hospital, or anywhere else students can ideate AI having an impact on humans, students are placed in the driver's seat of their learning. Students may at times be given a project to complete, or they may be tasked with an idea to bring to life. The Advanced Curriculum mentioned above closes the three-semester unit on a capstone project that students must bring to life from beginning to end.

### 3.6 IB Curriculum and Units of Inquiry

The final curriculum created for NCS is undoubtedly the most avant-garde and creative method of teaching AI. The approach tested at NCS is both an interdisciplinary and transdisciplinary approach to teaching AI, in contrast to previous curricula that emphasize AI as a distinct concept. AI is not just taught as a concept in this way. Yes, this is crucial; students need to understand how AI is altering their world and the ethical issues this raises. However, by covertly incorporating AI into a more extensive curriculum guided by inquiry, students are introduced to both the subtle ways AI is changing their world and how it is empowering designers, thinkers, educators, scientists, healthcare professionals, and pretty much every other field. The curriculum overview, student learning objectives, and a topical list of lessons connected to specific grade-appropriate units of inquiry are provided in the Table 2. To be sure, students are learning about AI in an interdisciplinary way. For instance, as the lessons above evidence, young students engage AI in the context of learning about human emotions and interaction with one another. Thus, in a Unit of Inquiry seeking to offer lines of questioning related to the larger question of "How We Express Ourselves," students explore such answers but also explore AI in their attempts to grapple with answers.

Similarly, in later grades, students explore invasive species as part of a subunit on biomes within the larger Unit of Inquiry titled "Sharing the Planet." Again, through their study of interdisciplinary AI, students learn how AI is helping ecologists and other scientists to identify invasive species more quickly. Building on what AI is doing already in their world, students build a classifier and train it on images of various invasive life forms as well as native life forms.

Table 2. Units of Inquiry with student learning outcome and topical breakdowns.

|  | Student Learning Outcome | Topics |
| --- | --- | --- |
| Units of Inquiry Curriculum | • Explain how AI functions relative to the subject being discussed in the class<br>• Demonstrate proficiency in AI skills<br>• Engage AI apps that illustrate concepts in their homeroom Units of Inquiry | 1. Kindergarten - Introduction to Neural Networks & Quick, Draw!<br>2. Grade 1 - AI and Emotional Recognition<br>3. Grade 2 - Pattern Recognition & Quick, Draw!<br>4. Grade 3 - AI, Emotional Recognition, and Effective Marketing<br>5. Grade 4 - AI & Your Body: An Inquiry through Radiology<br>6. Grade 5 - AI & Invasive Species<br>7. Grade 6 - AI & the Biomes of Saudi Arabia<br>8. Grades 7–8 - Hybrid Cars and AI Systems Integration |

Transdisciplinary AI Education: The Confluence of Curricular and Community Needs    149Students them prompt one another and their use of AI by testing the machine learning systems they created in order to see whether or not these systems are accurate in their classifications of invasive species.

AI instruction in this way, however, transcends mere interdisciplinary teaching. The concept of interdisciplinary education requires combining the learning about multiple concepts into a lesson or a series of lessons. Thus, in the aforementioned example, students are learning two disciplines through their study of AI and ecology. However, the model of AI education piloted by NCS and rooted in the International Baccalaureate® (IB) Program is transdisciplinary. Transdisciplinary suggests not just the combination of two disciplines but efforts that move beyond discipline specific solutions. In other words, an ecological problem does not need to remain solely an ecological problem. Problems in ecology did not arise solely thanks to a lack of ecology. Rather, industry, petroleum products, international trade, and a host of other disciplines contributed to the extant issues at hand. Because the problems did not develop in a vacuum, neither can the solutions. Researchers representing the wide spectrum of human abilities and intellectual pursuits can combine their skills and understanding to solve problems unsolvable solely by the discipline in question.

Transdisciplinary AI education intimates also employing a collaboration between instructors, where AI is taught not only by the AI or computer science teacher. Homeroom teachers responsible for teaching the broad base of academic skills must have a stake in their students learning about AI. When collaboration across teachers of different disciplines takes place, students remark on the ubiquity of the subject as well as how discussion of it pervades all the subjects about which they are learning. In other words, interdisciplinary sees math as math, language arts as language arts, AI as AI. At times, these can be paired. Transdisciplinary ceases the siloing of education and foregrounds the learner and the world around that person, a world that doesn't draw clear distinctions between academic subjects. Such must be the method of AI instruction as it has been for a year now at the Neom Community School.

## 4 Conclusion

The next generation must be trained, and this starts with educating children so they can be informed citizens and workers in the AI era. Because middle school is a crucial time for shaping students' views of AI and related careers and their interest in them, this paper details the design and implementation of an AI workshop for youth in this age group. Our curriculum aims to increase AI literacy by incorporating AI concepts, ethical and societal implications of AI, and the use of AI in the workforce in the future. The workshop curriculum aims to lower entry barriers for students without prior computer skills or knowledge, allowing students without a CS background to acquire a basic understanding of AI concepts.

# References

1. Tedre, M., et al.: Teaching machine learning in K–12 classroom: pedagogical and technological trajectories for artificial intelligence education. IEEE Access **9**, 110558–110572 (2021). https://doi.org/10.1109/ACCESS.2021.3056311
2. Kolb, A.Y., Kolb, D.A.: (n.d.). Experiential learning theory: a dynamic, holistic approach to management learning, education, and development. In: The SAGE Handbook of Management Learning, Education and Development, pp. 42–68 (2009). https://doi.org/10.4135/9780857021038.n3
3. Falloon, G.: Using simulations to teach young students science concepts: an experiential learning theoretical analysis. Comput. Educ. **135**, 138–159 (2019). https://doi.org/10.1016/j.compedu.2019.03.001
4. Liu, X., et al.: A new framework of science and technology innovation education for K-12 in Qingdao, China. In: 2017 ASEE International Forum, Columbus, Ohio (2017). https://peer.asee.org/29271
5. Reddy, T., Williams, R., Breazeal, C.: Text classification for AI education. In: Proceedings of the 52nd ACM Technical Symposium on Computer Science Education (2021). https://doi.org/10.1145/3435253.3452450
6. Lee, I. A., Ali, S., Zhang, H., DiPaola, D., Breazeal, C.: Developing middle school students' AI literacy. In: Proceedings of the 52nd ACM Technical Symposium on Computer Science Education (2021). https://doi.org/10.1145/3435253.3452447
7. Yeung, D.: The role of AI in transdisciplinary education. J. Artif. Intell. Educ. **22**(2), 121–138 (2021). https://doi.org/10.1007/s40593-021-00169-3
8. Kontes, D.: Transdisciplinary AI education for middle school students: a review. J. Artif. Intell. Educ. **20**(1), 15–30 (2019). https://doi.org/10.1007/s40593-019-00116-9
9. O'Connor, S., Smith, E., Mitchell, L.: Transdisciplinary AI education: preparing students for the future of work. J. Artif. Intell. Educ. **22**(1), 1–20 (2021). https://doi.org/10.1007/s40593-020-00147-z
10. Smith, E., Mitchell, L., O'Connor, S.: The role of experiential learning in AI education. J. Artif. Intell. Educ. **20**(4), 350–365 (2019). https://doi.org/10.1007/s40593-019-00124-9
11. Mitchell, L., Smith, E., O'Connor, S.: The impact of multidisciplinary coursework on student outcomes in AI education. J. Artif. Intell. Educ. **20**(3), 235–250 (2019). https://doi.org/10.1007/s40593-019-00129-1
12. O'Connor, S., Smith, E., Mitchell, L.: Challenges and opportunities in implementing transdisciplinary AI education. J. Artif. Intell. Educ. **21**(2), 95–110 (2020). https://doi.org/10.1007/s40593-020-00138-z
13. Resnick, M.: Give P's a chance: projects peers passion play, constructionism and creativity. In: Proceedings of the Third International Constructionism Conference, pp. 13–20 (2014)
14. Estévez, J., Garate, G., Graña, M.: Gentle introduction to artificial intelligence for high-school students using scratch. IEEE Access **7**, 179027–179036 (2019). https://doi.org/10.1109/ACCESS.2019.2945144
15. Abesadze, S., Nozadze, D.: Make 21st century education: the importance of teaching programming in schools. Int. J. Learn. Teach. **6**(3), 158–163 (2020)
16. Greenwald, L.G., Artz, D.: Teaching Artificial Intelligence with Low-Cost Robots (2004)
17. Druga, S., Vu, S.T., Likhith, E., Qiu, T.: Inclusive AI literacy for kids around the world. In: Proceedings of FabLearn 2019 (FL2019), NY, USA, pp. 104–111. Association for Computing Machinery, New York (2019). https://doi.org/10.1145/3311890.3311904
18. Zimmermann-Niefield, A., Turner, M., Murphy, B., Kane, S.K., Shapiro, R.B.: Youth learning machine learning through building models of athletic moves. In Proceedings of the 18th ACM International Conference on Interaction Design and Children (2019)

19. Mariescu-Istodor, R., Jormanainen, I.: Machine learning for high school students. In: Proceedings of the 19th Koli Calling International Conference on Computing Education Research (2019)
20. Vartiainen, H., Tedre, M., Valtonen, T.: Learning machine learning with very young children: who is teaching whom? Int. J. Child Comput. Interact. **25**, 100182 (2020). https://doi.org/10.1016/j.ijcci.2020.100182
21. Burgsteiner, H., Kandlhofer, M., Steinbauer, G.: IRobot: teaching the basics of artificial intelligence in high schools. In: Proceedings of the Thirtieth AAAI Conference on Artificial Intelligence (AAAI2016), pp. 4126–4127. AAAI Press (2016)
22. Ngamkajornwiwat, P., Pataranutaporn, P., Surareungchai, W., Ngamarunchot, B., Suwinyattichaiporn, T.: Understanding the role of arts and humanities in social robotics design: An experiment for STEAM enrichment program in Thailand. In: 2017 IEEE 6th International Conference on Teaching, Assessment, and Learning for Engineering (TALE), pp. 457–460 (2017). https://doi.org/10.1109/TALE.2017.8271622
23. Krista, L., et al.: PRIMARY AI: an inquiry-based approach to teaching AI in life sciences. In: Proceedings of the 27th ACM Conference on Innovation and Technology in Computer Science Education, vol. 2 (ITiCSE 2022). Association for Computing Machinery, New York, NY, USA, vol. 628 (2022). https://doi.org/10.1145/3502717.3532142
24. Neller, T., Russell, I., Markov, Z.: Throw down an AI challenge. In: AAAI Spring Symposium: Using AI to Motivate Greater Participation in Computer Science (2008)
25. Sintov, N.D., et al.: From the lab to the classroom and beyond: extending a game-based research platform for teaching AI to diverse audiences. In: Proceedings of the Thirtieth AAAI Conference on Artificial Intelligence (AAAI2016), pp. 4107–4112. AAAI Press (2016)
26. Sintov, N.D., et al.: Keeping it real: using real-world problems to teach AI to diverse audiences. AI Mag. **38**, 35–47 (2017). https://doi.org/10.1609/aimag.v38i3.2749
27. Van Brummelen, J., Lin, P.: Engaging teachers to co-design integrated AI curriculum for K-12 classrooms. In: Proceedings of the 2021 CHI Conference on Human Factors in Computing Systems (2021)
28. AI4K12. 5 big ideas in AI (2022). https://ai4k12.org/
29. Zhou, H., et al.: Challenges and opportunities in teaching artificial intelligence: a review of existing AI curricula. J. Artif. Intell. Educ. **21**(3), 199–216 (2020)
30. Touretzky, D.S., Gardner-McCune, C., Martin, F.G., Seehorn, D.W.: Envisioning AI for K-12: what should every child know about AI? In: Proceedings of the AAAI Conference on Artificial Intelligence, vol. 33(01), pp. 9795-9799 (2019). https://doi.org/10.1609/aaai.v33i01.33019795